\documentclass[twocolumn,prb,superscriptaddress]{revtex4}
\usepackage{graphicx}
\usepackage{color}\usepackage{enumerate}
\usepackage{amsmath}\usepackage{amssymb}\usepackage{stmaryrd}
\usepackage{multirow}
\usepackage{float}
\usepackage{longtable,booktabs}
\usepackage{subfigure}
\usepackage{dsfont}
\usepackage{amssymb}
\usepackage[pdftex,colorlinks=true,linkcolor=blue,citecolor=blue]{hyperref}

\begin{document}
\title{Many-Body Electric Multipole Operators in Extended Systems}

\author{William A. Wheeler}
\affiliation{Department of Materials Science and Engineering, University of Illinois at Urbana-Champaign, IL 61801, USA}
\author{Lucas K. Wagner}
\affiliation{Department of Physics and Institute for Condensed Matter Theory, University of Illinois at Urbana-Champaign, IL 61801, USA}
\author{Taylor L. Hughes}
\affiliation{Department of Physics and Institute for Condensed Matter Theory, University of Illinois at Urbana-Champaign, IL 61801, USA}

\begin{abstract}
The quantum mechanical position operators, and their products, are not well-defined in systems obeying periodic boundary conditions. Here we extend the work of Resta\cite{resta1998}, who developed a formalism to calculate the electronic polarization as an expectation value of a many-body operator, to include higher multipole moments, e.g., quadrupole and octupole. We define $n$-th order multipole operators whose expectation values can be used to calculate the $n$-th multipole moment when all of the lower moments are vanishing (modulo a quantum). We show that changes in our operators are tied to flows of $n-1$-st multipole currents, and encode the adiabatic evolution of the system in the presence of an $n-1$-st gradient of the electric field. Finally, we test our operators on a set of tightbinding models to show that they correctly determine the phase diagrams of topological quadrupole and octupole models, capture an adiabatic quadrupole pump, and distinguish a bulk quadrupole moment from other mechanisms that generate corner charges.
\end{abstract}

\maketitle

The modern theory of electric polarization in crystals has had a wide-ranging impact over the past 25 years\cite{king-smith1993,vanderbilt1993,resta1994,ortiz1994,resta2007,vanderbilt2018}. Aspects of the theory are not only useful for predictions of observable properties of real materials, but are tightly connected to a variety of topological insulator phenomena. For example, insulating crystals having a symmetry that would naively forbid the existence of a dipole moment, may instead have non-vanishing, but quantized, polarization if they are in a topological insulator phase\cite{zak1989,qi2008,hughes2011inversion,turner2012}. Furthermore, by tracking changes in polarization in an adiabatic cycle, one can observe quantized charge pumping characterized by a Chern number topological invariant\cite{thouless1983}.  More recently, the theory of quantized electronic polarization in topological insulators was extended to account for higher electric multipole moments\cite{benalcazar2016,benalcazar2017}. Therein it was shown that two and three dimensional, non-interacting topological crystalline phases of matter can support quantized electric quadrupole or octupole moments. 

The electric polarization is calculated by the Berry phase of the electronic energy bands over non-contractible cycles in the Brillouin zone\cite{king-smith1993,vanderbilt1993}. Similarly, Ref. \onlinecite{benalcazar2016} adapts this Berry phase formalism to describe the hierarchy of electric multipole moments.  While the Berry phase calculations for higher multipole moments can also be used on non-topological insulators with moments outside the quantized regime, the formulae are not directly applicable to many-body interacting systems. For polarization, one can consider the Berry phase of a many-body ground state parameterized by twisted periodic boundary conditions\cite{ortiz1994,souza2000}, but it is not obvious how this approach can be practically extended to the nested Berry phase (Wilson loop) approach to calculating higher multipole moments.  

In this letter we instead provide a formulation of the quadrupole and octupole moments in terms of ground-state  expectation values of many-body operators, analogous to Resta's formulation of the charge polarization\cite{resta1998}. We define many-body multipole operators, and discuss the connection between these operators and the flow of adiabatic multipole currents. In addition, we test our operators for several explicit model systems. We show that our formulae precisely capture the phase diagrams of the quadrupole and octupole models from Ref. \onlinecite{benalcazar2016}.
Furthermore, we use our formula to track the changes in quadrupole moment during the higher order pumping process in Ref. \onlinecite{benalcazar2017}, and we show that the results match the expected phenomena exactly. As an important comparison, we go on to test our formula on a model with corner charges but no bulk quadrupole moment. We show that our operator does not detect any difference between the phases of this model with and without corner charges, as we expect since there is no change in the bulk quadrupole moment, despite the presence of corner charges. 
%

\section{Motivation}

Let us begin by recounting the calculation of the electric polarization using a many-body expectation value\cite{resta1998}. Consider a translationally invariant, crystalline insulator with periodic boundary conditions. It is well known that the many-body position operator $\hat{{\bf{X}}}=\sum_{n}\hat{{\bf{x}}}_n,\; (n=1, \ldots N_e)$ cannot be naively used in the calculation of the electric polarization in extended (and/or periodic) systems. In such systems the operator is not well-defined because it can transform a state in the Hilbert space to a state outside the Hilbert space, e.g., it can take a normalizable state to a non-normalizable state, or one obeying periodic boundary conditions to one that violates the boundary conditions.  Instead, the operator \begin{equation}\label{eq:restasoperator} \hat{\mathcal{U}}_j=\exp\left[\frac{2\pi i  \hat{X}^j}{L_j}\right]\end{equation} can be employed in calculations of the polarization of a many-body ground state $\vert \Phi_0\rangle$, i.e., $P^j=\frac{eL_j}{2\pi \mathcal{V}}{\rm{Im}}\log\langle \Phi_0\vert \hat{\mathcal{U}}_j\vert \Phi_0\rangle\equiv \frac{eL_j}{2\pi \mathcal{V}}{\rm{Im}}\log z^{(P)}_j,$ where $\mathcal{V}$ is the volume of the system. 
This formula for $P^j$ approximates a derivative with respect to the many-body momentum by a finite difference \cite{souza2000}, so it is only strictly true in the thermodynamic limit $L_j\rightarrow\infty$.
For a non-degenerate, insulating ground state one finds $\hat{\mathcal{U}}_j\vert \Phi_0\rangle=e^{i\gamma_j}\vert\Phi_0\rangle +O(1/L_j)$ as the thermodynamic limit is approached\cite{resta1998}, i.e., $\vert \Phi_0\rangle$ becomes an eigenstate of $\hat{\mathcal{U}}_j$ in the thermodynamic limit\footnote{In taking  the thermodynamic limit one must take $L_j\to\infty$ first before the transverse directions.} (we recount this result briefly below) with a polarization given by $P^{j}=\frac{e\gamma_j}{2\pi}\frac{L_j}{\mathcal{V}}.$
The polarization has an ambiguity from the choice of the branch of the $\log,$ often referred to as the quantum of polarization, i.e., $\gamma_j\equiv \gamma_j+2\pi n$.

Following this line of reasoning we can consider higher electric multipole moments such as the quadrupole $\hat{q}^{ij}=\sum_{n}\hat{x}^{i}_{n}\hat{x}^{j}_n$ or octopole $\hat{o}^{ijk}=\sum_{n}\hat{x}^{i}_{n}\hat{x}^{j}_n\hat{x}^{k}_n.$  These operators are problematic in periodic systems for the same reasons as the many-body position operator, but we can also consider exponentiated versions. Let ${\bf{b}}_{a}$ be the set of reciprocal lattice vectors satisfying ${\bf{b}}_{a}\cdot {\bf{a}}_{b}=\delta_{ab}$ for the primitive unit vectors ${\bf{a}}_{b}.$ Then we can consider the operators
\begin{eqnarray}
\label{eq:UQ}\hat{\mathcal{U}}^{Q}_{ab}&=&\exp\left[\frac{2\pi i b_{a}^{i}\hat{q}^{ij}b_{b}^{j}}{N_a N_b} \right]\\ \label{eq:UO}
\hat{\mathcal{U}}^{O}_{abc}&=&\exp\left[\frac{2\pi i\hat{o}^{ijk}b_{a}^{i}b_{b}^{j}b_{c}^{k}}{N_a N_b N_c} \right]
\end{eqnarray} where $N_{a}$ is the number of unit cells in the $a$-th lattice direction. The bulk moments are then defined as
\begin{eqnarray}
Q^{ab}&=&\frac{e L_a L_b}{2\pi \mathcal{V}}{\rm{Im}}\log \langle \Phi_0\vert \hat{\mathcal{U}}^{Q}_{ab}\vert \Phi_0\rangle\\
O^{abc}&=&\frac{e L_a L_b L_c}{2\pi \mathcal{V}}{\rm{Im}}\log \langle \Phi_0\vert \hat{\mathcal{U}}^{O}_{abc}\vert \Phi_0\rangle
\end{eqnarray} where we have left the crucial step of taking the thermodynamic limit implicit in these formulas. For the majority of this article we will direct our focus to the quadrupole operator, and the octupole case can be studied essentially mutatis mutandis.


\section{Moments as integrated current}

To confirm that our definition of the quadrupole moment matches the expected physical observables we will generalize the arguments from Ref. \onlinecite{resta1998}. Consider a generic many-body Hamiltonian
\begin{equation}
H=\sum_{n=1}^{N_{elec}}\sum_{i=x,y,z} \frac{(\hat{p}_{n}^i)^2}{2m} +V({\bf{\hat{X}}}).
\end{equation}\noindent Let us focus on the unitary transformation $\hat{U}_{\lambda}(\alpha)=\exp\left(i \alpha \lambda({\bf{\hat{X}}})\right)$, for an arbitrary real differentiable function $\lambda(\bf{\hat{X}})$, that acts on each momentum operator as
\begin{eqnarray}
\hat{U}_{\lambda}(\alpha)\hat{p}^{i}_n\hat{U}_{\lambda}^{\dagger}(\alpha)=\hat{p}^{i}_n -\hbar\alpha \frac{\partial \lambda({\bf{\hat{X}}})}{\partial \hat{x}^{i}_n}.
\end{eqnarray} Now we define $H(\alpha)=\hat{U}_{\lambda}(\alpha)H\hat{U}_{\lambda}^{\dagger}(\alpha),$ noting that the difference between $H(0)$ and $H(\alpha)$ is the presence of an electromagnetic vector potential acting on each electron, e.g., $A^{i}({\bf{\hat{x}}}_n)=-\frac{\hbar\alpha}{e} \frac{\partial \lambda({\bf{\hat{X}}})}{\partial \hat{x}^{i}_n}$ for the $n$-th electron. Assuming that the ground state of $H(0)$, $\vert \Psi_0\rangle$, is  always non-degenerate, then the relation $H(\alpha)U_{\lambda}(\alpha)\vert \Psi_0\rangle=E_0 U_{\lambda}(\alpha)\vert \Psi_0\rangle$ holds. 



Let us take the particular case where $\lambda({\bf{\hat{X}}})=\hat{X}^1,$ which will allow us to reproduce Resta's argument. This generates the constant vector potential $A^1=-\frac{\hbar \alpha}{e}, A^2=A^3=0$. To have $U_{\hat{X}^1}(\alpha)\vert\Psi_0\rangle$ satisfy periodic boundary conditions, we can choose $\alpha = 2\pi/L_1.$ When approaching the thermodynamic limit, $L_1$ is large and we can treat $\alpha$ as a small parameter. Thus we can treat  $U_{\hat{X}^1}(2\pi/L_1)\vert \Psi_0\rangle,$ which is an eigenstate of $H(2\pi/L_1),$ as a perturbed version of the initial ground state $\vert \Psi_0\rangle,$ and hence we can expand $U_{\hat{X}^1}(2\pi/L_1)\vert \Psi_0\rangle$ in terms of the eigenstates of $H(0).$ The perturbation term in the Hamiltonian is $H'=-\frac{2\pi\hbar}{mL_x}\sum_{n=1}^{N_{elec}}\hat{p}_{n}^{1},$ and the leading order correction to the unperturbed ground state yields
\begin{eqnarray}\label{eq:upeigenstate}
&&U_{\hat{X}^1}(2\pi/L_1)\vert \Psi_0\rangle \approx\nonumber\\&& e^{i\gamma_{p1}}\left(\vert \Psi_0\rangle -\frac{2\pi\hbar}{mL_1}\sum_{j\neq 0}\vert\Psi_j\rangle \frac{\langle \Psi_j\vert \hat{\mathcal{P}}^1\vert \Psi_0\rangle}{E_0-E_j}\right)
\end{eqnarray}\noindent where $\vert \Psi_j\rangle$ are the excited states of $H(0)$ with energies $E_j,$ and $\hat{\mathcal{P}}^i$ is the many-body momentum operator of all the electrons. In addition to the usual form, we have allowed for a phase factor $\gamma_{p1}$ which determines the $1$-component of the electric polarization\cite{resta1998}. As we stated earlier, we now see explicitly that in the thermodynamic limit $U_{\hat{X}^1}(2\pi/L_1)\vert \Psi_0\rangle =e^{i\gamma_{p1}}\vert \Psi_0\rangle.$ To connect to the physical polarization we can show that the time-derivative of this polarization definition matches the physically expected result of an electric current.  Indeed Ref. \onlinecite{resta1998} shows from the perturbation theory calculation above that in 1D
\begin{eqnarray}
\frac{d P^1}{dt}\approx \frac{ie\hbar}{mL_1}\sum_{j\neq 0}\langle \dot{\Psi}_0\vert \Psi_j\rangle \frac{\langle \Psi_j\vert \hat{\mathcal{P}^1}\vert \Psi_0\rangle}{E_0-E_j}+{\textrm{c.c}},
\end{eqnarray} which is the adiabatic electric charge current\cite{thouless1983}.

Now consider the quadrupole moment $\hat{Q}^{12}$. We will consider a unitary transformation $\hat{U}_\lambda(\alpha)$ with $\lambda({\bf{\hat{X}}})=\sum_{n=1}^{N_{elec}}\hat{x}^1_n \hat{x}^2_n$ and with $\alpha=2\pi/L_1 L_2$. Using the same arguments as the case for polarization, we can use perturbation theory to calculate \begin{eqnarray}\label{eq:quadphaseperturb}
&&U_{\hat{\mathcal{Q}}^{12}}(2\pi/L_1L_2)\vert \Psi_0\rangle \approx\nonumber\\&&e^{i\gamma_{q12}}\left(\vert \Psi_0\rangle
-\frac{2\pi\hbar}{L_1L_2}\sum_{j\neq 0}\vert\Psi_j\rangle \frac{\langle \Psi_j\vert \hat{\mathcal{J}}_{D}^{12}\vert \Psi_0\rangle}{E_0-E_j}\right)
\end{eqnarray}\noindent where $\hat{\mathcal{J}}_{D}^{12}=\tfrac{1}{m}\sum_{n=1}^{N_{elec}} (\hat{p}_{n}^1 \hat{x}_{n}^2+\hat{p}_{n}^2 \hat{x}_{n}^1)$ is the \emph{dipole-current} operator. Note that the velocity $d\hat{x}^{i}_{n}/dt=(i/\hbar)[H,\hat{x}^{i}_{n}]=p^{i}_{n}/m$ is like the time derivative of the dipole operator, while the time derivative of the quadrupole operator is $d(\hat{x}^1_n \hat{x}^{2}_n)/dt=(i/\hbar)[H,\hat{x}^1_n \hat{x}^{2}_n]=(1/m)(\hat{p}^{1}_n \hat{x}^{2}_n+\hat{p}^{2}_n \hat{x}^{1}_n),$ hence our notion of a dipole current operator. From this result we can further calculate 
\begin{eqnarray}\label{eq:dipolecurrent}
\frac{d Q^{12}}{dt}\approx \frac{ie\hbar}{L_1L_2}\sum_{j\neq 0}\langle \dot{\Psi}_0\vert \Psi_j\rangle \frac{\langle \Psi_j\vert \hat{\mathcal{J}}_{D}^{12}\vert \Psi_0\rangle}{E_0-E_j}+{\textrm{c.c}}\nonumber\\
\end{eqnarray}\noindent which relates the time derivative of our definition of the quadrupole to a dipole-current as we would physically expect. We note that the above expression for $\hat{\mathcal{J}}_{D}^{12}$ (and thus Eqs. \ref{eq:quadphaseperturb} and \ref{eq:dipolecurrent}) cannot be evaluated in a periodic system in its current form because it contains position operators.
Nonetheless, the notion of a dipole current is useful for understanding the origin of a bulk quadrupole moment as the result of a process that rearranges charge moments, even in extended systems.
The problem with periodicity is reminiscent of the polarization operator ${\bf\hat{X}}$, which cannot be evaluated directly in a periodic system, but can be computed from the operator in Eq. \ref{eq:restasoperator}.
Finding an alternate expression for  $\hat{\mathcal{J}}_{D}^{12}$ that is compatible with periodic boundary conditions is an interesting direction for future work.

\section{Adiabatic evolution}

For an intuitive understanding we can characterize the action of our operators in the language of adiabatic evolution. The action of $\hat{U}_{\hat{X}^1}(2\pi/L_1)$ on a state can be treated as adiabatic evolution from a system with a vanishing vector potential to one with $A^{1}=-h/eL_1$.  The process can be accomplished through a time-dependent vector potential of the form $A^1=-\frac{ht}{eL_1 T},$ and which is $0$ for $t<0,$ and $-h/eL_1$ for $t>T.$ The process inserts one magnetic flux quantum into the periodic cycle in the $1$-direction. From the Faraday effect, this is equivalent to turning on a uniform electric field in the $1$-direction during the time interval $t \in [0, T].$ We are interested in the evolution of the non-degenerate ground-state of a gapped, neutral insulator, and so the presence of the uniform electric field will only activate the additional phase factor $\gamma_{p1}=-\hbar^{-1}\int_{0}^{T} {\bf{d}}\cdot {\bf{E}}\,dt$ in the adiabatic limit, where ${\bf{d}}$ is the total dipole moment of the material. Besides the weak, time-varying vector potential, we assume the insulator is otherwise static so that ${\bf{d}}$ is time-independent. Hence, we find the phase (c.f. Eq. \ref{eq:upeigenstate})
\begin{eqnarray}
\gamma_{p1}&=&\hbar^{-1}d^{1}\int_{0}^{T}\partial_t A^{1}dt=d^{1}\cdot(A^1(T)-A^1(0))\nonumber\\
&=&-2\pi d^{1}/eL_x
\end{eqnarray}\noindent which exactly produces the right value of  $P^{1}=\frac{e}{2\pi}{\rm{Im}}\log\langle \Phi_0\vert \hat U_{\hat{X}^1}(2\pi/L_1)\vert \Phi_0\rangle$ in the thermodynamic limit.  

The essential idea in the previous paragraph is that, to detect a polarization in the material, we adiabatically turn on a small electric field and track how the phase of the ground state responds during this process. For the quadrupole,  instead of turning on a uniform electric field, we turn on a uniform electric field gradient. The quadrupolar operator $U_{\hat{\mathcal{Q}}^{12}}(2\pi/L_1L_2)$ can be interpreted as adiabatically evolving our ground state from a vanishing vector potential at $t=0$ to the vector potential $A^{i}=-\frac{h\sigma^{ij}x^j}{eL_1 L_2}$ at $t=T.$ This process can be carried out using a time-dependent vector potential 
\begin{equation}
A^{i}=-\frac{h t \sigma^{ij}x^j}{eL_1 L_2 T}\;\;\;\; i, j =1 ,2,
\end{equation}\noindent where $\sigma^{12}=\sigma^{21}=1, \sigma^{11}=\sigma^{22}=0,$ and $T$ is large. This vector potential represents a constant electric field gradient. Thus, a natural way to interpret the proposed quadrupole operator is as an evolution process where a small, uniform electric field gradient is turned on for a finite amount of time. During this time, the ground state will develop the phase factor $\gamma_{q12}$ shown in Eq. \ref{eq:quadphaseperturb}, in the thermodynamic limit. Since we are applying the electric field gradient in a neutral, unpolarized insulator, and the $Q^{12}$ quadrupole moment couples to an electric field gradient $\tfrac{1}{2}(\partial_1 E^2+\partial_2 E^1)$, its contribution to the phase will be $\gamma_{q12}=-\tfrac{1}{2\hbar}\int_{0}^{T}(q^{12}\partial_1 E^2+q^{21}\partial_2 E^1).$ We assume that the quadrupole moment of our system is static so we can simplify this expression to  
\begin{eqnarray}
\gamma_{q12}&=&\frac{q^{12}}{2\hbar}\int_{0}^{T}(\partial_1 \partial_t A^{2}+\partial_2\partial_t A^{1})dt\nonumber\\
&=&\frac{q^{12}}{2\hbar}(\partial_1 (A^2(T)-A^{2}(0))+\partial_2(A^{1}(T)-A^{1}(0))\nonumber\\&=&-\frac{2\pi q^{12}}{eL_1 L_2},
\end{eqnarray} which confirms our operator definition of the quadrupole moment. 

\section{Practical evaluation}

After having shown that $\hat{\mathcal{U}}^{Q}_{ab}$ can be used to determine dipole currents in the bulk, and that it has a physical interpretation via adiabatic evolution in an electric field gradient, we now turn to the practical evaluation of the expectation value of this operator. 
To determine if $\hat{\mathcal{U}}^{Q}_{ab}$ can have a well-defined, non-vanishing expectation value $z^{(Q)}_{ab}$ in periodic systems we need to evaluate its transformation properties under translations. 
Let us recall the argument from Ref. \onlinecite{aligia1999} that shows that $z_{j}^{(P)}$ is non-vanishing only when the expectation value of $\hat{\mathcal{U}}_j$ transforms in the trivial representation of the translation group. Assuming translation invariance,  the ground state $\vert \Phi_0\rangle$ is an eigenstate of the translation operator carrying many-body momentum ${\bf{K}}.$ We find
\begin{eqnarray}&&\langle \Phi_0\vert \hat{\mathcal{U}}^{P}_{a}\vert \Phi_0\rangle=\langle \Phi_0\vert\hat{T}_{{\bf{a}}_c}^{-1}\hat{T}_{{\bf{a}}_c}\hat{\mathcal{U}}^{P}_{a}\hat{T}_{{\bf{a}}_c}^{-1}\hat{T}_{{\bf{a}}_c}\vert \Phi_0\rangle\nonumber\\&=&e^{i{\bf{K}}\cdot {\bf{a}}_c}e^{-i{\bf{K}}\cdot {\bf{a}}_c}\exp(2\pi i\delta_{ac}n_{0c})\langle \Phi_0\vert \hat{\mathcal{U}}^{P}_{a}\vert \Phi_0\rangle
\end{eqnarray}  where we translated all the electrons in the $c$-th primitive direction using $\hat{T}_{{\bf{a}}_c},$ and assumed $\vert\Phi_0\rangle$ is an eigenstate of the total number operator. We see that if $n_{0a}=N_e/N_a$ is an integer for $N_a$ unit cells in the $a$-th direction, then the expectation value does not have to vanish. We can relax $n_{0a}$ to be rational with a suitable modification of $\hat{\mathcal{U}}_a$\cite{aligia1999}.

Repeating for the quadrupole we find 
\begin{eqnarray}
&&\langle \Phi_0\vert \hat{\mathcal{U}}^{Q}_{ab}\vert \Phi_0\rangle=
\exp\left[2\pi i n_{0ab}\delta_{ac}\delta_{bc}\right]\nonumber\\
&\times& \langle\Phi_0\vert \hat{\mathcal{U}}^{Q}_{ab} \exp\left[2\pi i \hat{X}^j(b_{a}^{j}\delta_{bc}+\delta_{ac}b_{b}^{j})/N_a N_b\right]\vert \Phi_0\rangle\nonumber\\
&=& e^{i\Gamma}\langle \Phi_0\vert \hat{\mathcal{U}}^{Q}_{ab}\vert \Phi_0\rangle+O(1/L)
\end{eqnarray}\noindent where $n_{0ab}=N_e/N_a N_b$, is assumed to be an integer, and we have applied the results of Ref. \onlinecite{resta1998} to evaluate $e^{2\pi i \hat{X}^j(b_{a}^{j}\delta_{bc}+\delta_{ac}b_{b}^{j})/N_a N_b}\vert\Phi_0\rangle=e^{i\Gamma}\vert\Phi_0\rangle+O(1/L).$ 
Hence, in the thermodynamic limit we find the relation $\langle \Phi_0\vert \hat{\mathcal{U}}^{Q}_{ab}\vert \Phi_0\rangle=e^{i\Gamma}\langle \Phi_0\vert \hat{\mathcal{U}}^{Q}_{ab}\vert \Phi_0\rangle,$ and $z^{(Q)}_{ab}$ must therefore vanish unless $\Gamma=2\pi p$ for some integer $p.$ The phase factor $\Gamma$ is different than the phase generated by translations of the polarization operator $\hat{\mathcal{U}}^{P}_{a}$, which depends on the particle number $n_{0a}$. Instead, $\Gamma$ depends on the polarization $\hat{X}^j$.
Specifying that the planar filling factor $n_{0ab}$ is an integer (or rational fraction for suitably generalized $\hat{\mathcal{U}}^{Q}_{ab}$\cite{aligia1999}) will not be enough in this case, and one must also specify that the polarization vanishes (up to a quantum).

However, even after satisfying this constraint, the expectation value of $\hat{\mathcal{U}}^{Q}_{ab}$ is not invariant under translations in general.
Constraining the polarization to vanish (which should ensure $\Gamma=2\pi p$) will not fix the issue in finite-sized systems due to the non-vanishing \emph{fluctuations} of the dipole moment. The dipole fluctuations are a measure of the Wannier function localization length, or alternatively, the non-vanishing correlation length controlled by the insulating gap. When the dipole fluctuations are non-vanishing, the state $\vert \Phi_0 \rangle$ is not an eigenstate of the $\hat{X}^j$ operator, so even with no average dipole, evaluating the expectation value of the quadrupole operator can be problematic.  If we have single particle orbitals in a zero-correlation length limit where Wannier functions are point-localized, then the dipole fluctuations vanish and the many-body ground state will be an exact eigenstate of the $\hat{\mathcal{U}}^{P}_a$ operators. For this situation the expectation value of $\hat{\mathcal{U}}^{Q}_{ab}$ is well-defined even in finite-size as long as the polarization vanishes, and the particle filling is integer. Indeed, for many-body ground states that are an exact eigenstate of $\hat{\mathcal{U}}^{P}_a$ the magnitude of the expectation value of $\hat{\mathcal{U}}^{Q}_{ab}$ will tend to unity in the thermodynamic limit.  To evaluate the expectation value of the quadrupole operator more generally we must be more careful.\footnote{Similar problems would occur for the polarization operator if the ground state was not an eigenstate of the number operator, but instead had fluctuations in particle number.}  The effect of dipole fluctuations on $\hat{\mathcal{U}}^{Q}_{ab}$ is discussed further in Appendix \ref{sec:scaling_magnitude}.


In this work we replace the quadrupole operator by an approximation: we evaluate the expectation value of $\hat{\mathcal{U}}_{ab}^{Q}$ on a finite supercell of size $L_x\times L_y$, then extend the truncated operator periodically outside the supercell. We then evaluate the expectation value of this periodic operator. The error introduced by this approximation depends on the size of the supercell we use,  but improves as the ratio of the correlation/localization length and the characteristic length of the supercell goes to zero. From our analytic evaluation of the expectation value of $\hat{\mathcal{U}}_{ab}^{Q}$ for Gaussian charge configurations in Appendix \ref{sec:scaling_magnitude}, we expect that the magnitude of the expectation value will approach a constant in a thermodynamic limit where both $N_x$ and $N_y$ are taken to infinity together. This constant will not generically be unity, and it will depend on the size of the dipole fluctuations in the $x$ and $y$ directions as well as the aspect ratio $N_x/N_y$. In this limit the phase factor that determines the quadrupole moment converges to the correct value as well. Discussions of finite size error and more details about the effects of dipole fluctuations are shown in Appendices \ref{sec:finite_size} and \ref{sec:scaling_magnitude}. We will leave a systematic study of this approximation and alternatives to future work.



\section{Example Calculations}
\subsection{Multipole Models with Mirror or $C_4$ Symmetry} 
\begin{figure}
\centering
\includegraphics[width=0.7\columnwidth]{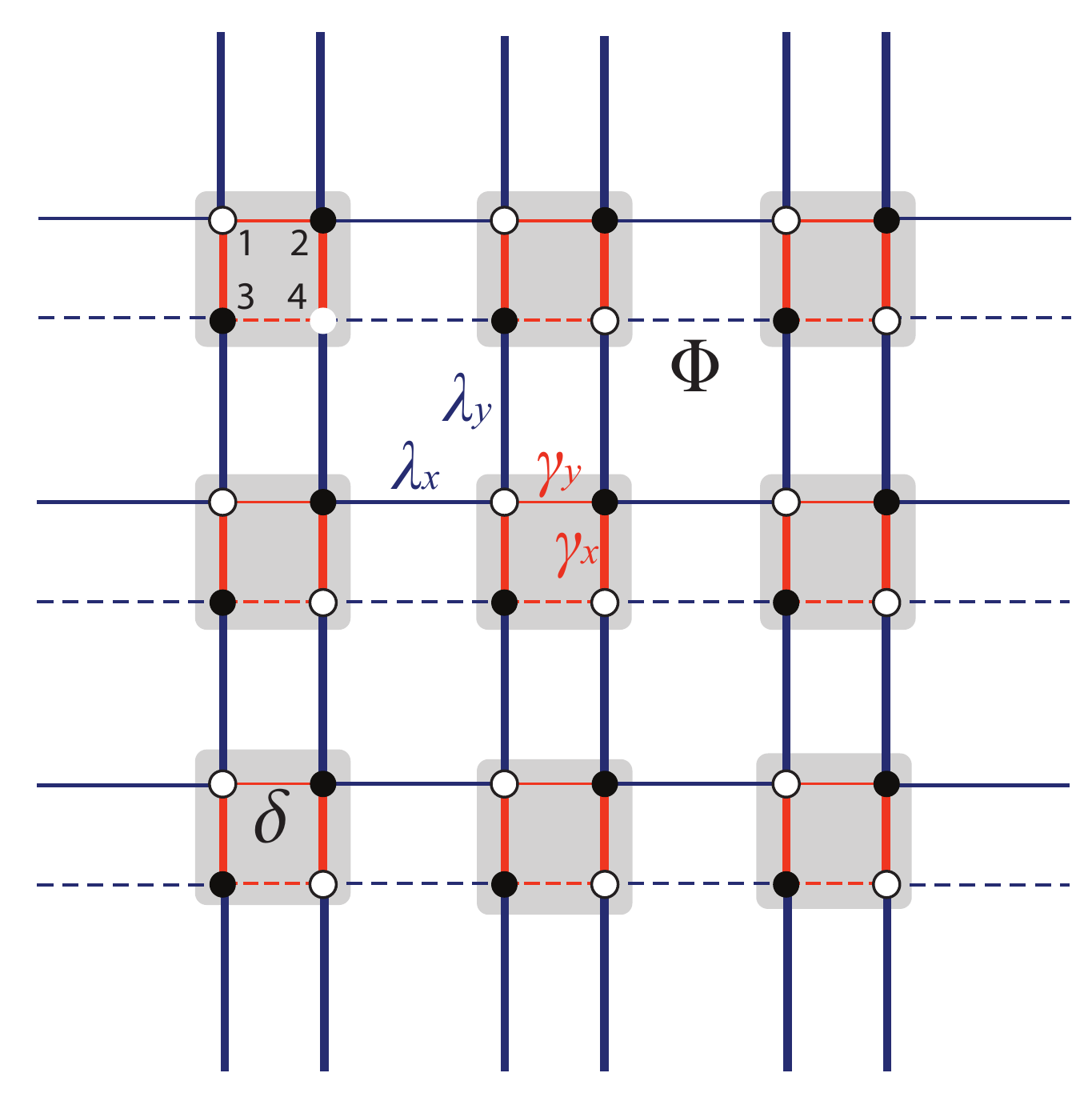}
\caption{Tightbinding model having four spinless orbitals per cell with inter-cell tunneling $\lambda_{x,y},$ intra-cell tunneling $\gamma_{x,y},$ and onsite potential $\delta$ that takes opposite signs on the filled and empty circles. There is a flux $\phi$ between each unit cell and within each unit cell in a gauge where the dotted lines have a relative phase factor of $e^{i\Phi}.$}
\label{fig:TBmodel}
\end{figure}

Now we provide some examples in which we calculate the multipole moments using our operators in tightbinding models. Our first example will be focused on the insulating phases of a particular tightbinding model\cite{benalcazar2016,benalcazar2017} on a square lattice with four degrees of freedom per unit cell, which, for simplicity, we will treat as spinless orbitals (we also provide calculations for a related octupole model\cite{benalcazar2016}, and in the subsequent subsection we provide calculations for another tightbinding model with a non-vanishing quadrupole moment, but with $C_4 T$ symmetry\cite{schindler2018}). The unit cell basis and tunneling terms are illustrated in Fig. \ref{fig:TBmodel}, and the Bloch Hamiltonian for a system with periodic boundary conditions is given by
\begin{eqnarray}
&&H({\bf{k}},\Phi)=\nonumber\\&&\left(\begin{array}{cccc}\delta &\Gamma_x(k_x)&\Gamma_y(k_y)&0\\
\Gamma_x(-k_x)&-\delta&0&\Gamma_y(k_y)\\
\Gamma_y(-k_y)&0&-\delta&e^{i\Phi}\Gamma_x(k_x)\\
0&\Gamma_y(-k_y)&e^{-i\Phi}\Gamma_x(-k_x)&\delta\end{array}\right)\nonumber\\\label{eq:TB}
\end{eqnarray}\noindent where $\Gamma_x(k_x)=\gamma_x+\lambda_x e^{ik_x}, \Gamma_y(k_y)=\gamma_y+\lambda_y e^{ik_y},$ $\delta, \gamma_{x/y}, \lambda_{x/y}$ are real parameters representing onsite energies, intra-cell tunneling, and inter-cell tunneling respectively, and $\Phi$ is the flux in each (intra- and inter-cell) plaquette in our chosen gauge. 

\begin{figure}
\centering
\includegraphics[width=\columnwidth]{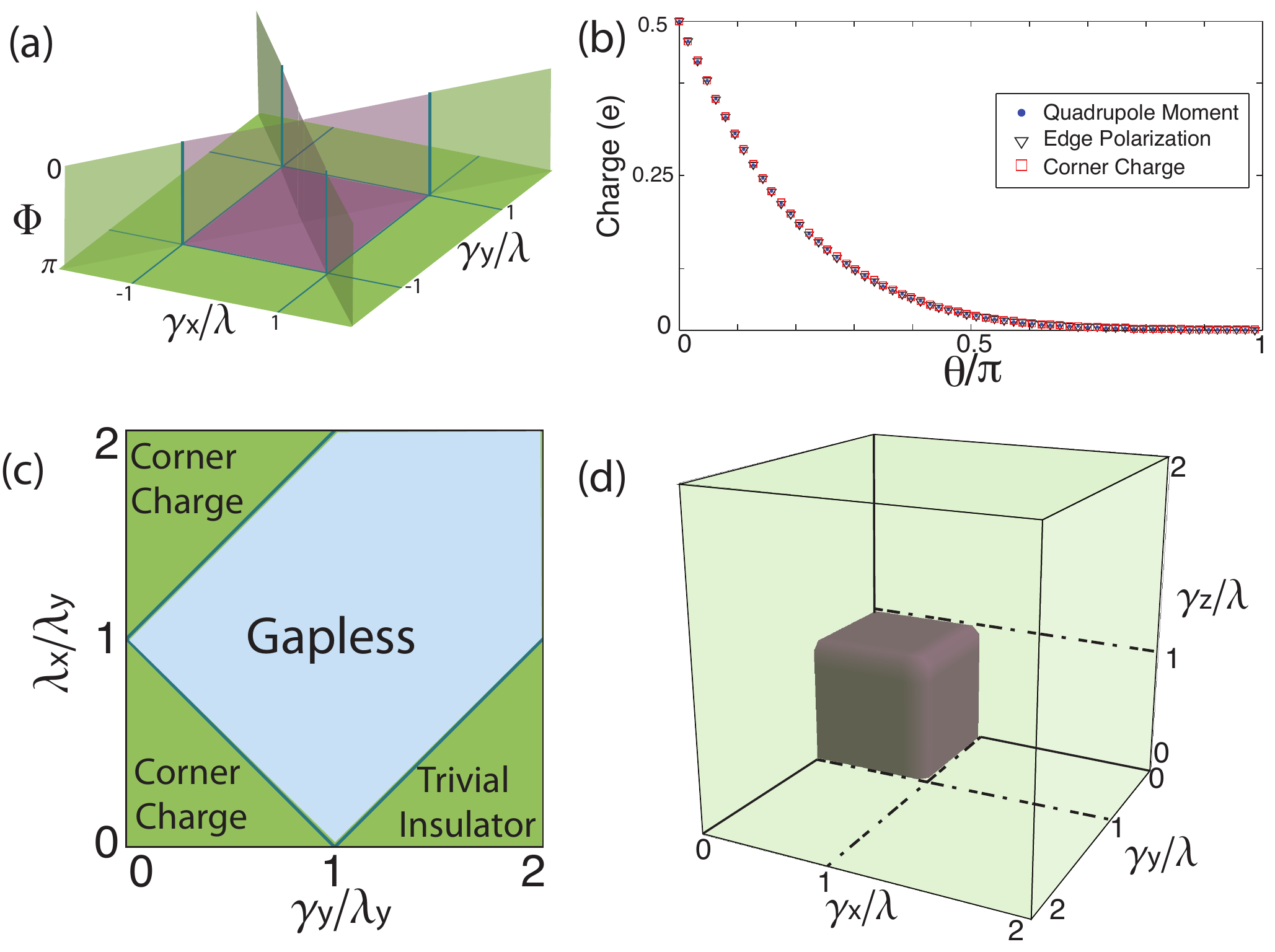}
\caption{(a)Phase diagram for model in Eq.~\ref{eq:TB} with $L_x=L_y=40$ as a function of $(\gamma_x/\lambda,\gamma_y/\lambda,\Phi).$ The green and purple regions have a values of $Q^{xy}$ that differ by $e/2,$ as calculated using Eq.~\ref{eq:UQ}. Dependence on $\Phi$ is only shown along the diagonal lines that maintain $C_4$ symmetry. (b) The evolution of the quadrupole moment (Eq.~\ref{eq:UQ}) during a pumping process as compared with the corner charge and edge polarization. All three match exactly, as expected. (c) Phase diagram for Eq.~\ref{eq:TB} with $\Phi=0$ as a function of $(\gamma_y/\lambda_y, \lambda_x/\lambda_y).$ Eq.~\ref{eq:UQ} does not distinguish two different phases that have corner charge and a trivial insulator with no corner charge. This is as expected as in this regime the model has a vanishing quadrupole moment even in the phases with corner charge/modes.\cite{benalcazar2017} (d) Phase diagram for an octupole model\cite{benalcazar2016} with $L_x=L_y=L_z=10$ as a function of $(\gamma_x/\lambda,\gamma_y/\lambda,\gamma_z/\lambda).$ The green and purple regions have values of $O^{xyz}$ that differ by $e/2$ as determined by Eq.~\ref{eq:UO}.}
\label{fig:quadphase}
\end{figure}

This model can be tuned to a variety of insulator phases, and we will consider several test cases. For case (i) we explore this model in the quadrupolar phase protected by $M_x$ and $M_y$ mirror symmetries where $\Phi=\pi,$ $\delta=0,$ and we fix $\lambda_x=\lambda_y\equiv\lambda$ to plot a phase diagram as a function of $\gamma_x/\lambda$ and $\gamma_y/\lambda$ in the interval $[-2, 2].$ We find that the calculation of the quadrupole moment using our operator (see Fig.~\ref{fig:quadphase}a), matches the expected phase diagram from Ref. \onlinecite{benalcazar2016} where the system has a quantized quadrupole moment with magnitude $Q^{xy}=e/2$ when $\gamma_x/\lambda, \gamma_y/\lambda$ are both in the interval $[-1, 1].$ 

Let us make some technical remarks about this calculation. Although difficult to see by eye, there is some deviation from the expected phase diagram very close to the phase boundaries where the correlation length is increasing without bound, but this deviation is just a finite-size effect as it decreases rapidly with the size of the supercell. Additionally, while the phase of the operator $\hat{\mathcal{U}}_{ab}^Q$ yields the quadrupole moment of the ground state in the thermodynamic limit, it can only be extracted if the magnitude of the expectation value is non-vanishing. According to Eq.~\ref{eq:quadphaseperturb}, one would expect its magnitude to approach one in the thermodynamic limit; however, we find that, for our choice of boundary conditions and our choice for the approach to the thermodynamic limit, the magnitude does not go to unity in the presence of dipole fluctuations. In Appendix~\ref{sec:finite_size}, we show the scaling of the magnitude of the operator as a function of system size for two points in the phase diagram shown in Fig.~\ref{fig:quadphase}a. Furthermore, we discuss the complications associated with dipole fluctuations in more detail in Appendix~\ref{sec:scaling_magnitude}.

For case (ii) we  consider a variation of the quadrupole phase protected by $C_4$ symmetry where we tune $\gamma_x=\gamma_y\equiv \gamma$, $\lambda_x=\lambda_y\equiv \lambda$ and allow for $\Phi \in [0, \pi].$ At $\Phi=0$ the model is gapless at half filling, but for any finite $\Phi$ in this interval we expect to find a quantized quadrupole of $Q^{xy}=e/2$ when $|\lambda|>|\gamma|.$ We show this calculation in the extended planes along the diagonals of the phase diagram in Fig.~\ref{fig:quadphase}a, and find that our operator correctly reproduces the phase diagram. 

For case (iii) we consider the adiabatic dipole pumping process introduced in Refs.~\onlinecite{benalcazar2016,benalcazar2017}  where the quadrupole moment is continuously tuned from the topological phase where $Q^{xy}=e/2$ to the trivial phase where $Q^{xy}=0.$ We can parameterize this pumping process by $\lambda_x=\lambda_y=1,$ $\delta\to -\sin(\theta(t)),$ and $\gamma=1/2(1-\cos(\theta(t))).$ We show the result in Fig.~\ref{fig:quadphase}b, in which we plot the result of our operator compared to calculations of the corner charge (for an open, square geometry) and the edge polarization on a cylinder, and we find they all match exactly as expected for a continuously varying bulk quadrupole moment $Q^{xy}.$

For case (iv) we tune the system to $\Phi=\delta=0.$ This model has $M_x$ and $M_y$ mirror symmetries so we expect the quadrupole moment to be quantized, though that does not mean it takes the non-trivial topological value. Indeed this model is not expected to have a bulk quadrupole moment despite having insulating phases with fractional $e/2$ corner charge, since this charge is associated with edge polarization rather than a bulk quadrupole moment. Here we show a phase diagram for fixed $\lambda_y=1, \gamma_x=0,$ where we vary $\lambda_x$ and $\gamma_y.$ We show that our formula, while yielding a quantized value for $Q^{xy},$ does not distinguish between the insulating phases with and without fractional corner charge. This is a clear and important indication that our invariant is only sensitive to the bulk quadrupole moment, and is not just testing for the possible existence of corner charge alone. 

Finally, in  Fig.~\ref{fig:quadphase}d we show a calculation for the octupole moment in the phase diagram of the mirror-symmetric octupole model of Ref.~\onlinecite{benalcazar2016} where we have fixed the inter-cell hopping $\lambda$ and varied the intra-cell hoppings $\gamma_{x,y,z}$ in the interval $[0,2].$ We see that the phase diagram matches the expected result with a topological octupole moment $O^{xyz}=e/2$ when $\vert \gamma_i/\lambda\vert<1$ for all $i.$

\begin{figure}
\centering
\includegraphics[width=0.75\columnwidth]{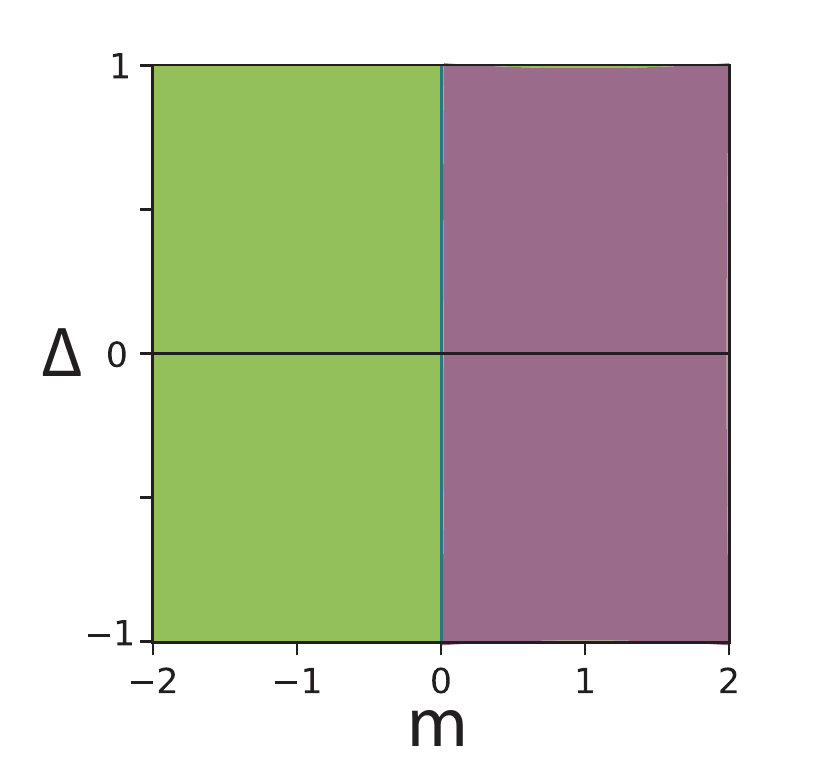}
\caption{Phase diagram for model in Eq.~\ref{eq:qsh} with $L_x=L_y=40$ as a function of $(m,\Delta).$ The green and purple regions have values of $Q^{xy}$ that differ by $e/2$ according to Eq.~\ref{eq:UQ}. The data along the line $\Delta=0$ is not shown because the values of $Q^{xy}$ are highly fluctuating, and the quadrupole moment is not well-defined since the edges are gapless.}
\label{fig:qshphase}
\end{figure}

\subsection{Quadrupole model with $C_4 T$ Symmetry}\label{app:qsh}
In this subsection we perform a calculation of the quadrupole moment in a 2D version of the chiral hinge-insulator model presented in Ref.~\onlinecite{schindler2018}. This model has a quantized quadrupole moment protected by the product of rotation and time-reversal symmetry $C_4 T.$ The Bloch Hamiltonian we consider is
\begin{eqnarray}\label{eq:qsh}
H({\bf{k}})&=&\sin k_x \Gamma^1+\sin k_y\Gamma^2\nonumber+(2-m-\cos k_x -\cos k_y)\Gamma^0\nonumber\\&+&\Delta (\cos k_x-\cos k_y)\Gamma^3    
\end{eqnarray}\noindent where $\Gamma^0=\tau^z\otimes\mathbb{I},\Gamma^1=\tau^x\otimes\sigma^x, \Gamma^2=\tau^x\otimes\sigma^y,\Gamma^3=\tau^x\otimes\sigma^z.$ When $\Delta=0$ this model has both $C_4$ and time-reversal symmetry $T,$ and is a model representing a 2D time-reversal invariant topological insulator with helical edge states when, e.g., $0<m<2$. When $\Delta\neq 0$ the system breaks both $C_4$ and $T$ but preserves the product $C_4 T.$ The non-vanishing $\Delta$ acts to gap the helical edge states and generate corner modes for an open square geometry that respects global $C_4$ symmetry. Thus we expect to find an ill-defined quadrupole moment when $\Delta=0,$ and a non-vanishing topological quadrupole moment $Q^{xy}=e/2$ when $0<m<2, \Delta\neq 0.$ When $m<0, \Delta\neq0$ we expect to find a well-defined, but vanishing quadrupole moment. We show the calculated phase diagram as a function of $m$ and $\Delta$ in Fig.~\ref{fig:qshphase}. We note that we look at a range of $\Delta$ that is small enough not to destroy the bulk topology and we see that our operator correctly reproduces the phase diagram. 

\section{Connection to recent work}
There have been several recent papers discussing the evaluation of the quadrupole operators on which we can comment\cite{gil2018,ono2019}.
Kang et. al.\cite{gil2018} simultaneously proposed the same definitions for the quadrupole and octupole operators as we do in Eqs.~2-5. 
Similar to our argument in Eq.~16, they investigate the invariance of the operator $\hat{\mathcal{U}}_{xy}$ under translations, and conclude that the polarization must be zero for the operator to be well-defined.
We have both pointed out that the ground state is not an eigenstate of polarization $\hat{\mathcal{U}}_j$, which is what complicates the evaluation of $\hat{\mathcal{U}}_{xy}.$
They point out a violation of translation invariance of $\mathcal{O}(1/E_{\rm gap})$, inversely proportional to the excitation gap:
	\[\langle \hat{\mathcal{U}}_{xy}' \rangle = \langle \hat{\mathcal{U}}_{xy} \rangle \langle \hat{\mathcal{U}}_y \rangle + \mathcal{O}\left(\frac{1}{E_{\rm gap}}\right),\]
	where $ \hat{\mathcal{U}}_{xy}'$ is the operator $ \hat{\mathcal{U}}_{xy}$ translated in the $x$ direction by $x \rightarrow x + L_x$. We describe this same error as being a result of dipole fluctuations, which, for non-interacting systems,  measure the Wannier localization length.
For single-particle orbitals, the error disappears in the limit that the Wannier functions are $\delta$-function localized, equivalent to their limit of $E_{\rm gap} \rightarrow \infty$. We drew a connection between dipole fluctuations and decreasing magnitude of $\langle \hat{\mathcal{U}}_{xy} \rangle,$ which is a signature of the problem with translation invariance.
In contrast to their discussion, we point out that zero polarization is not sufficient to have a well-defined quadrupole in finite-sized periodic systems.
As such, for systems with dipole fluctuations, evaluating this operator in PBCs necessarily uses an approximate form; we described one approximation that we used in our calculation, which is similar to that used in Ref. \onlinecite{gil2018}.

To confirm the definitions of the quadrupole operator $\hat{\mathcal{U}}_{xy}$ as a reliable estimator of the response to electric field gradients, they also give an argument similar to our adiabatic evolution discussion, but in a more field-theoretical approach.



In response to these proposals, Ono et. al.\cite{ono2019} comment on some problems that arise when applying the quadrupole operator $\hat{\mathcal{U}}_{xy}$ to certain tight-binding models, that were not considered in our work\cite{ono2019}. They find that their evaluation of the quadrupole operator has more serious issues than those discussed here, namely that the result seems to depend non-trivially on the correlation length in an insulator phase when the quadrupole moment should nominally be quantized. They also comment that their results can depend on the parity of the number of lattice sites, but as we show in Appendix A, C this latter issue has a simple origin, and similar effects can appear even for the polarization, and even for classical point-charge configurations.

In the models in which we evaluated the quadrupole moment we did not find any dependence on the correlation length besides typical finite-size effects, so it might be useful to comment on the distinctions between the models we considered and one of their models, though fully resolving the discrepancy will require further research. One of their models is a $C_4$ invariant system with four occupied bands that form a Wannier representable obstructed atomic limit. The electrons in the model sit at maximal Wyckoff positions $1a, 1b, 2c$, i.e., one at the center, one at the corner, and two in the middle of the edges of the unit cell. One possibly important distinction is that this model does not have gapped Wannier bands. While the need for gapped Wannier bands was emphasized in Ref. \onlinecite{benalcazar2017}, it has not been proven that this is a generic requirement to define the quadrupole moment (though a gapped, neutral edge \emph{is} required). A second distinction is that in the models we consider, one can always relate the configurations with a non-vanishing quadrupole to a configuration representing a trivial onsite limit through a continuous deformation that preserves a symmetry under which the \emph{dipole moment does not change}. Thus, in our models one can clearly calculate a \emph{change} in quadrupole moment when going from the trivial atomic limit to a non-vanishing quadrupole configuration. In their model, such an interpolation does not seem to exist, so it is not clear how to compare the quadrupole moment in their obstructed atomic limit to the trivial atomic limit. While again, this may not be a requirement to define a quadrupole model, it does seem like a natural consideration.

Ref. \onlinecite{ono2019} also makes a note regarding second-order contributions to the perturbative expansion of the magnitude $|\langle \Phi_0 | \hat {\mathcal{U}}_x | \Phi_0 \rangle|$, and comments that Resta's formula for polarization is not guaranteed to work, i.e., the magnitude may go to zero in higher dimensions. 
This is a well-known fact for polarization. Indeed to obtain a nonzero magnitude of the expectation value of $\hat{\mathcal{U}}_x$ in higher dimensions one must take the correct thermodynamic limit, namely, where the direction parallel to the polarization component of interest goes to infinity first.
We also find that, for periodic boundary conditions, the quadrupole magnitude $|\langle \hat{\mathcal{U}}_{xy} \rangle|$ goes to zero in the thermodynamic limit if dipole fluctuations are finite, seemingly no matter what thermodynamic limit one uses. Despite this we find that this issue does not affect the consistency of the phase in the tight-binding calculations we performed in finite size systems. Based on the calculations in Appendix C, we expect that if we truncate the system, as indicated by our approximation scheme, then the magnitude $|\langle \hat{\mathcal{U}}_{xy} \rangle|$ can approach a finite (though likely not unity) value if we take the limit $L_x=L_y\to \infty,$ even in the presence of dipole fluctuations.

In summary, Ono et. al. raise some interesting questions about the evaluation of the many-body quadrupole operator and about what constraints are necessary to enforce in order to have a well-defined quadrupole moment in a crystalline system. The issues they have pointed out will be valuable in specifying careful definitions and treating boundary effects correctly for detecting higher order multipole moments in bulk systems.


\section{Conclusion}

In conclusion, we have proposed definitions for many-body operators whose expectation values determine the quadrupole and octupole moments of insulators. 
We showed that the change in quadrupole moment corresponds to dipole currents in the material.
We proposed a method to evaluate the quadrupole; however, the method is not entirely satisfactory because it suffers from significant finite size errors related to the fluctuations of the dipole moment, even when the total dipole moment is vanishing. However, we showed that on several nontrivial tightbinding models the operators do indeed capture the bulk properties, and can be used to evaluate topological indices.
The operator can be evaluated using many-body wave functions as well as using tightbinding wave functions, so it can be used to generate new many-body topological indices for systems with interactions and/or disorder.

{\bf{Note:}} During the preparation of this manuscript we became aware of an independent overlapping work by B. Kang, K. Shiozaki, and G. Y. Cho\cite{gil2018}. We thank them for discussions and for coordinating submission.

\acknowledgements{We thank Jahan Claes, Matthew Foulkes, and Gerardo Ortiz for useful conversations. LKW would like to thank the Simons Foundation Collaboration on the many electron problem for support. TLH thanks the US National Science Foundation under grant DMR 1351895-CAR, and the MRSEC program under NSF Award Number DMR-1720633 (SuperSEED) for support.}

\appendix

\section{Constraints on Charge Configurations for the Quadrupole and Octupole Operators}
\label{sec:appconstraint}

In this Appendix we will evaluate the conditions under which one will generically find a non-vanishing expectation value of multipole operators if the electron orbitals are point-charge localized.
For example, in a limit where all dipole fluctuations vanish, we expect that the expectation values of the quadrupole operators are well-behaved in extended systems. More precisely, to evaluate the quadrupole moment we want to consider a many-body state that is an exact eigenstate of the $\mathcal{U}_x$ and $\mathcal{U}_y$ operators. This is not necessarily an unphysical situtation, as we know it will be approximately true as we approach the thermodynamic limit. That is, if we consider our insulating system to be large compared to the characteristic size of individual Wannier functions of the occupied bands, then we can treat the charge distribution as a collection of point charges located at the Wannier centers that are periodically repeated throughout the crystal lattice\cite{vanderbilt1993}. For a many-body state represented by a product of single-particle orbitals, vanishing dipole fluctuations imply that the orbitals are $\delta$-function localized, so we will consider the expectation values of the multipole operators in such a state. Even in this situation, there are constraints under which the expectation values of the quadrupole or octupole operators are well-defined in the thermodynamic limit, i.e, when they effectively transform in the trivial representation of the translation group, for such point-charge product states. We will determine those conditions now.

For simplicity, let us consider our electron configuration to be a rectangular lattice of point charges and focus on the quadrupole moment $Q^{xy}$ associated to the operator
\begin{equation}
\hat{\mathcal{U}}^{Q}_{xy}=\exp\left(\frac{2\pi i\sum_{j} x_j y_j}{L_xL_y}\right).\label{eq:uqxy}
\end{equation} Now let us transform this operator by a translation ${\bf{R}}=h_1a_x \hat{x}+h_2a_y\hat{y}$ where $h_1, h_2$ are integers. We find that
\begin{widetext}
\begin{eqnarray}
&&T_{{\bf{R}}}\hat{\mathcal{U}}^{Q}_{xy}T_{{\bf{R}}}^{-1}=\hat{\mathcal{U}}^{Q}_{xy}\exp\left[2\pi i\sum_j h_1 h_2/N_x N_y\right]\exp\left[2\pi i h_1\sum_{j}y_j/N_xL_y\right] \exp\left[2\pi i h_2\sum_{j}x_j/L_x N_y\right]\nonumber\\
&&= \hat{\mathcal{U}}^{Q}_{xy}\exp\left[2\pi i\sum_{{\bf{R}}}\sum_{\alpha=1}^{\nu} h_1 h_2/N_x N_y\right]\exp\left[2\pi i h_1\sum_{{\bf{R}}}\sum_{\alpha=1}^{\nu}y_{{\bf{R}},\alpha}/N_xL_y\right] \exp\left[2\pi i h_2\sum_{{\bf{R}}}\sum_{\alpha=1}^{\nu}x_{{\bf{R}},\alpha}/L_x N_y\right]\nonumber\\&&=\hat{\mathcal{U}}^{Q}_{xy}\exp\left[2\pi i\nu h_1 h_2\right]\exp\left[2\pi i h_1\sum_{n_2}\sum_{\alpha=1}^{\nu}(\bar{y}_{\alpha}+n_2 a_y)/L_y\right] \exp\left[2\pi i h_2\sum_{n_1}\sum_{\alpha=1}^{\nu}(\bar{x}_{\alpha}+n_1 a_x)/L_x\right]\nonumber\\
&&=\hat{\mathcal{U}}^{Q}_{xy}\exp\left[2\pi i\nu h_1 h_2\right]\exp\left[\frac{2\pi i h_1}{a_y}\sum_{\alpha=1}^{\nu}\bar{y}_{\alpha}\right] \exp\left[\frac{2\pi i h_2}{a_x}\sum_{\alpha=1}^{\nu}\bar{x}_{\alpha}\right]\exp\left[\pi i h_1\nu (N_y+1)\right] \exp\left[\pi i h_2\nu (N_x+1)\right],
\end{eqnarray}\noindent where the sum over $\alpha$ runs over all the electrons in a single unit cell and $\nu$ is the electron filling. \end{widetext}

Ultimately we want this operator to be invariant under any lattice translation of all the electrons, and the strongest constraints arise from taking $h_1, h_2$ to be the smallest non-zero integers, e.g., $h_1=h_2=1.$ For this choice we find the constraint\begin{widetext}
\begin{equation}
\exp\left[2\pi i\nu\right]\exp\left[\frac{2\pi i}{a_y}\sum_{\alpha=1}^{\nu}\bar{y}_{\alpha}\right] \exp\left[\frac{2\pi i}{a_x}\sum_{\alpha=1}^{\nu}\bar{x}_{\alpha}\right]\exp\left[\pi i \nu (N_y+N_y+2)\right]=1. 
\end{equation}\end{widetext} To satisfy this constraint in a manner that is independent of the eveness/oddness of $N_x, N_y$ it is sufficient to choose $\nu=2\mathbb{Z}$ and the $x$ and $y$ components of the dipole moment per unit cell must be integer multiples of $a_x, a_y$ respectively. We can relax the constraint of $\nu$ being integer valued if we modify our starting operator Eq.~\ref{eq:uqxy} according to a prescription analogous to that presented in Ref.~\onlinecite{aligia1999}, but in that case we will still require $\nu = 2\mathbb{Q}.$ While we have derived these constraints assuming the charge distribution is point-like we expect this result to apply to more complicated distributions in the thermodynamic limit where the system size is much larger than the spatial extent of the Wannier functions that compose our insulating ground state. The coordinates $\bar{x}_\alpha, \bar{y}_\alpha$ will then be associated to the centers of the Wannier functions in each unit cell.

We can repeat this calculation for the octupole moment, e.g, $O^{xyz},$ and translating by a vector ${\bf{R}}=a_x \hat{x}+a_y \hat{y}+a_z \hat{z}.$ We find the constraint that the dipole and quadrupole moments per unit cell must be integer valued as well as the constraint
\begin{equation}
\exp\left[\frac{2\pi i\nu (N_x+1)(N_y+1)(N_z+1)}{4}    \right]=1.
\end{equation} In order for this constraint to be satisfied independently of the system size we thus need $\nu=4\mathbb{Z}$ for the octupole moment.


\section{Finite size effects of the quadrupole operator}
\label{sec:finite_size}

\begin{figure}
    \centering
    \includegraphics[width=\columnwidth]{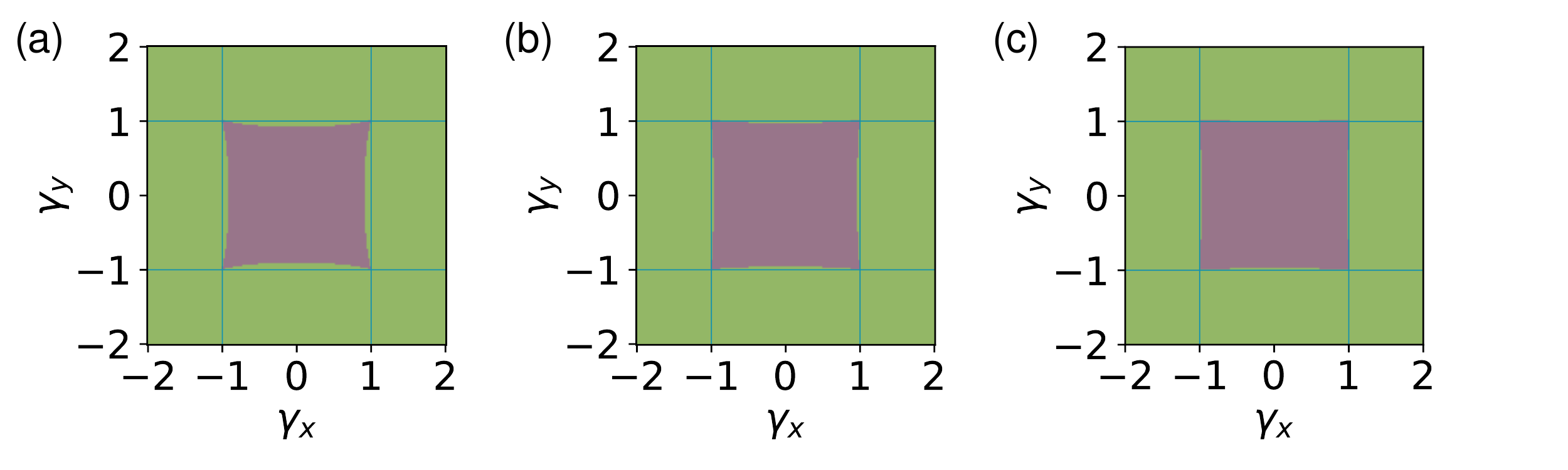}
    \caption{Phase diagrams in $\gamma_x$-$\gamma_y$ of ${\rm Im} \log \langle\Phi_0\vert \hat{\mathcal{U}}_{xy}^Q \vert\Phi_0\rangle$ for $N_x=N_y=10$ (a), $20$ (b), $40$ (c). Teal lines at $|\gamma_i|=1$ indicate the phase boundary of the model. Our operator gives a quadrupole moment of zero in the green regions and 0.5 in the purple regions. In larger systems, our operator's quadrupole phase more closely approaches the theoretical phase boundary.}
    \label{fig:finite_size_phase}
\end{figure}

The evaluation of the expectation value of the quadrupole operator $\hat{\mathcal{U}}^Q_{xy}$ (Eq.~\ref{eq:uqxy}) exhibits some challenges. We show how this manifests in our calculation as a function of system size. Fig.~\ref{fig:finite_size_phase} shows the $\gamma_x$-$\gamma_y$ phase diagram of the quantized, mirror-symmetric quadrupole model Eq. \ref{eq:TB} from Ref.~\onlinecite{benalcazar2017} for $L_x=L_y=10,\;20,\;40$. There is an error in a thin region near the phase boundary, where only one of $\gamma_x$ and $\gamma_y$ is close to (but less than) one. The expectation value predicts the wrong topological phase in this region; however, this region decreases rapidly with system size. In the thermodynamic limit, the error disappears. This type of error is due to the large correlation length that exceeds the system size for parameters near the phase boundaries.

\begin{figure}
    \centering
    \includegraphics[width=\columnwidth]{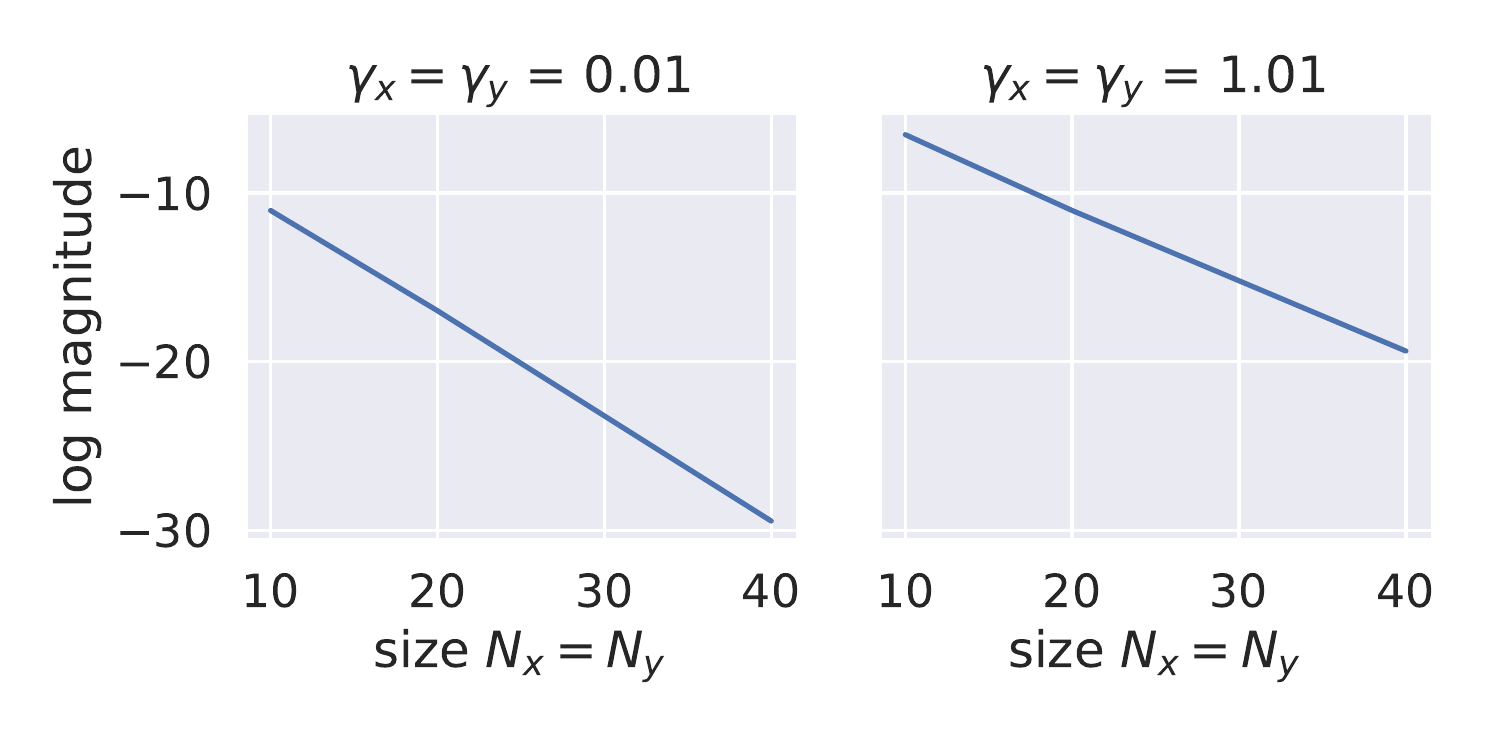}
    \caption{Log of the magnitude $\vert \langle\Phi_0\vert \hat{\mathcal{U}}_{xy}^Q \vert\Phi_0\rangle \vert$ at different points in the $\gamma_x$-$\gamma_y$ phase diagram. The magnitude goes to zero with increasing system size at all points in the phase diagram.}
    \label{fig:finite_size_mag}
\end{figure}

The expectation value of the operator $\hat{\mathcal{U}}^{Q}_{xy}$ has a phase and a magnitude. While the phase gives the quadrupole moment, the magnitude reflects its fluctuations. Fig.~\ref{fig:finite_size_mag} shows how the magnitude depends on our finite simulation size for selected points in the $\gamma_x$-$\gamma_y$ phase diagram shown in Fig.~\ref{fig:finite_size_phase}. 
The magnitudes are close to zero and decrease as the system gets larger.
This does not mean that the physical quantity is not meaningful; in comparison, for a 3D system, Resta's polarization operator also decreases to zero magnitude if $N_x=N_y=N_z\rightarrow \infty$. Exploring efficient ways to evaluate the expectation value for large systems is still an open challenge.
In Appendix~\ref{sec:scaling_magnitude} we will show how this magnitude decrease can be traced back to fluctuations in the dipole moments.

\section{Scaling of the Magnitude of the Quadrupole Expectation Value}\label{sec:scaling_magnitude}
In this Appendix we discuss the scaling properties of the expectation values of the dipole and quadrupole operators using a test configuration of continuum charge distributions arranged on a crystalline lattice. We will consider the cases of product states of $\delta$-function point charges as well as Gaussian charges, both product and determinant states.
In these systems, the dipole fluctuations cause the magnitude of the quadrupole operator to go to a limiting value which is not equal to one.

\subsection{Polarization}

For the calculation of the polarization let us consider a 1D system with a filling of one electron per unit cell at a position $x_0$ with respect to the origin of each unit cell. The electrons in our system are thus located at the positions $x_0+n$  in units of the lattice constant $a$ which we set to $a=1$, and where $n\in \mathbb{Z}.$  Now we can calculate the polarization via the expectation value of the operator $\hat{\mathcal{U}}_x = \exp\left[\frac{2\pi i \hat{X}}{L_x}\right]$ in the many-body ground state consisting of a tensor product of localized $\delta$-function charges. The polarization is given by
\begin{equation} P_x=\frac{e L_x}{2\pi \mathcal{V}}{\rm{Im}}\log\left[\langle \hat{\mathcal{U}}_x\rangle\right]
\end{equation} where $\mathcal{V}$ is the volume of the system, i.e., $L_x=N_x$ in 1D. After integrating over the real-space coordinates of each electron to generate the expectation value, we find the result
\begin{equation}\label{eq:polpoint}
P_x=ex_0+e\frac{N_x+1}{2},
\end{equation} which is only well-defined mod $e,$ so the second term either contributes $0$ or $1/2$ depending on the parity of $N_x.$  

Now let us take localized charges obeying a Gaussian distribution 
\begin{equation}
\psi_{x_0}(x)=\frac{1}{\sqrt{\pi^{1/2}\sigma_x}}\exp\left[-\frac{(x-x_0)^2}{2\sigma_{x}^2}\right].
\end{equation}
We find that, for a single charge at $x_0$
\begin{equation}
\langle \hat{\mathcal{U}}_x\rangle= \exp\left[\frac{2\pi i x_{0}}{N_x}\right]\exp\left[-\frac{\pi^2 \sigma_{x}^2}{N_x^2}\right].
\end{equation}\noindent We see that this result includes a phase factor encoding the position $x_0$ as well as an exponential factor with magnitude $\leq 1.$ The phase factor is identical to what the $\delta$-function charge distribution produces, and when all electrons across the lattice are considered it will exactly reproduce Eq. \ref{eq:polpoint} for the value of the polarization. On the other hand, if we have a number of electrons $N_e=N_x$ then the \emph{magnitude} of the expectation value will be $e^{-\pi^2\sigma_{x}^2/N_x}$ which will tend to unity as $N_x\to\infty$ as expected. Thus, this operator recovers unitarity in the thermodynamic limit.

Let us now illustrate an important subtlety by reconsidering the case of Gaussian charges in a 2D system. Suppose that we have one electron per unit cell on a square lattice, and that the electrons are centered at $(x_0, 0)+{\bf{R}}$ where ${\bf{R}}$ is the set of 2D lattice vectors. For a single-electron, which is Gaussian distributed in $x$ and $y$ we have
\begin{equation}
\langle \hat{\mathcal{U}}_x\rangle= \exp\left[\frac{2\pi i x_{0}}{N_x}\right]\exp\left[-\frac{\pi^2 \sigma_{x}^2}{N_x^2}\right]
\end{equation}\noindent which is the same as the 1D case above. The distinction between dimensions appears when we use the fact that $N_e=N_x N_y.$ After including all of the electrons we have
\begin{equation}
\vert\langle \hat{\mathcal{U}}_x\rangle\vert= \exp\left[-\frac{\pi^2 \sigma_{x}^2}{N_x^2}\right]^{N_x N_y}=\exp\left[-\frac{\pi^2 \sigma_{x}^2 N_y}{N_x}\right].
\end{equation} The important feature is that the magnitude of this operator approaches unity in a specific thermodynamic limit where we take $N_x\to\infty$ before $N_y,$ but it vanishes if we use the opposite order, and approaches a constant $<1$ if we take the limit while keeping a fixed aspect ratio $N_x/N_y$. Under the same conditions, but in 3D, we would find a factor $e^{-\pi^2\sigma_{x}^2 N_y N_z/N_x}.$ Thus, generically to keep the operator expectation value non-vanishing we must always take the thermodynamic limit in the direction of the polarization component of interest before letting the transverse directions approach infinity. 

\subsection{Quadrupole Moment}
Now let us consider the quadrupole moment for a 2D system on a square lattice. We can begin with the periodic arrangement of $\delta$-function point charges and calculate the expectation value of $\hat{\mathcal{U}}_{xy}= \exp\left[\frac{2\pi i \hat{Q}_{xy}}{L_x L_y}\right].$ We must be careful in this case to specify that the total dipole moment vanishes (up to an integer). We will satisfy this constraint by taking the system with two electrons per cell with coordinates $(x_0, y_0)$ and $(-x_0, -y_0)$ with respect to the origin of the cell, and we repeat these positions across the entire lattice. For this configuration, and for a lattice indexed from $1\dots N_x, 1\ldots N_y,$ we find
\begin{equation}
Q_{xy}=2e x_0 y_0 + \frac{e(N_x+1)(N_y+1)}{2},
\end{equation}\noindent which depends on the parity of $N_x$ and $N_y$ similarly to the polarization.

Now let us move on to charges with a Gaussian distribution described by
\begin{equation}
\psi_{(x_0, y_0)}(x)=\frac{1}{\sqrt{\pi\sigma_x\sigma_y}}\exp\left[-\frac{(x-x_0)^2}{2\sigma_{x}^2}-\frac{(y-y_0)^2}{2\sigma_y^{2}}\right].
\end{equation}
 In this case, the expectation value generated by a single electron with a Gaussian wave function centered at $(x_0, y_0)$ is
\begin{eqnarray}
&&\langle\hat{\mathcal{U}}_{xy}\rangle=\frac{2}{\sqrt{4+\alpha^2\sigma_x^2 \sigma_y^2}}\nonumber\\&&\times\exp\left[-\frac{\alpha^2 \sigma_{y}^2 x_0^2+\alpha^2 \sigma_{x}^2 y_{0}^2-4i\alpha x_0 y_0}{4+\alpha^2\sigma_{x}^2\sigma_{y}^2}\right]\nonumber\\&\equiv& \Lambda \exp\left[\frac{i\alpha x_0 y_0}{1+\alpha^2\sigma_{x}^2\sigma_{y}^2/4}\right]
\end{eqnarray}\noindent where $\alpha = 2\pi/L_x L_y.$ We can see from this result that both the magnitude \emph{and} the phase of this expectation value depend on the system size when $\sigma_x$ and/or $\sigma_y$ are non-vanishing. This is not the case for the polarization, where the imaginary phase is independent of the system size for a Gaussian distribution. Here the difference arises because we allow for fluctuations of the dipole moment, which is equivalent to having non-vanishing $\sigma_i.$ The dipole moment is vanishing on average, but has non-zero fluctuations that lead to this ambiguity in the phase factor. The same would likely occur for the polarization if we let the charge in each unit cell fluctuate, but we have implicitly assumed that it does not fluctuate. We can calculate the quadrupole moment density from the imaginary part of the $\log$ of this expression for the expectation value  to find
\begin{equation}
x_0 y_0\left(1-\frac{\pi^2\sigma_{x}^2\sigma_{y}^2}{L_x^2 L_y^2}+\ldots\right)
\end{equation}\noindent for a single electron. We find that this value converges to the correct result as $L_x, L_y\to \infty,$ i.e., when the system size is much larger than the dipole fluctuation lengths. When all of the electrons in the lattice are taken into account this will reproduce the value calculated for the $\delta$-function distribution.

Finally, let us consider the magnitude $\Lambda$ of this expectation value. We find a magnitude for a single electron to be
\begin{equation}
\Lambda=\frac{2}{\sqrt{4+\alpha^2\sigma_x^2 \sigma_y^2}}\exp\left[-\frac{\alpha^2 \sigma_{y}^2 x_0^2+\alpha^2 \sigma_{x}^2 y_{0}^2}{4+\alpha^2\sigma_{x}^2\sigma_{y}^2}\right].
\end{equation} We want to determine the value of this magnitude for a collection of electrons on the lattice. The magnitude for the full ground state is harder to calculate since it depends on the positions of the electrons, but for a product state including  two electrons per unit cell, i.e., $N_e=2N_xN_y,$ and ignoring contributions of the charge density that spreads from the last cells back into the first cells due to periodic boundary conditions,  we find\begin{widetext}
\begin{eqnarray}
&&\Lambda_{N_e}\approx\left(\frac{2}{\sqrt{4+\alpha^2\sigma_x^2 \sigma_y^2}}\right)^{2N_x N_y}\exp\left[-\frac{2\pi^2(\sigma_{y}^2 (N_x^2 /3 +x_0^2)+\sigma_{x}^2 ( (N_y^2 /3 +y_{0}^2) )}{N_x N_{y}+\pi^2 \sigma_x^2 \sigma_y^2/(N_x N_y)}\right]\nonumber\\&\approx&
\left(\frac{1}{\sqrt{1+\pi^2\sigma_x^2 \sigma_y^2/N_x^2 N_y^2}}\right)^{2N_x N_y}\exp\left[-\frac{(2/3)\pi^2(\sigma_{y}^2 N_x/ N_y+\sigma_{x}^2N_y/ N_x  )}{1+\pi^2 \sigma_x^2 \sigma_y^2/(N_x^2 N_y^2)}\right]
\end{eqnarray} where in the first approximation we have dropped the terms that shift $N_{x,y}$ by $1$ coming from a discrete sum over unit cells, and in the second approximation we have dropped the terms proportional to $x_0, y_0$ since they are coordinates within a unit cell and are less than $1$ in units of the lattice constant.  We find that as $N_x,N_y\to\infty$ that the first factor tends to unity:
\begin{eqnarray}
\left(\frac{1}{\sqrt{1+\pi^2\sigma_x^2 \sigma_y^2/N_x^2 N_y^2}}\right)^{2N_x N_y}\approx \left(1-\frac{\pi^2\sigma_x^2 \sigma_y^2}{2N_x^2 N_y^2}\right)^{2N_x N_y}\approx \exp\left(-\pi^2\sigma_x^2\sigma_y^2/N_x N_y\right)\rightarrow 1.
\end{eqnarray} If we make an analogy with the polarization, this factor has a dependence on the quadrupole fluctuations $\sigma_x^2 \sigma_y^2/4$ and tends to unity as long as those fluctuations are finite. 

Now let us consider the second factor 
\begin{equation}
\exp\left[-\frac{(2/3)\pi^2(\sigma_{y}^2 N_x/ N_y+\sigma_{x}^2N_y/ N_x  )}{1+\pi^2 \sigma_x^2 \sigma_y^2/(N_x^2 N_y^2)}\right].
\end{equation} This factor has interesting features: (i) the numerator has a dependence on the dipole fluctuations in the $x$ and $y$ directions, (ii) the denominator has a dependence on the quadrupole fluctuations. If dipole fluctuations in either direction vanish, then there is a consistent way to take the thermodynamic limit such that this term tends to unity. If the dipole fluctuations are non-vanishing in both directions then the best that one can do is to take $N_x=N_y\to \infty.$ In this case the factor tends to the finite value $e^{-2\pi^2(\sigma_y^2+\sigma_{x}^2)/3},$ which is non-vanishing, but not unity. We also note that if one takes $N_x$ or $N_y$ to infinity first then this factor generically tends to zero. This result seems to imply that even though $\hat{\mathcal{U}}_{xy}$ may not be strictly a well-defined operator in periodic systems with dipole fluctuations, its expectation value is still meaningful and can recover the correct results for the quadrupole moment. 
\end{widetext}

The above discussion considers a product state of Gaussian orbitals; however, the tightbinding calculations presented in the paper use a Slater determinant, so we also show a comparison with a determinant of the Gaussian orbitals.
An analytic calculation of the determinant state is impractical, so we compared the numerical results of the determinant and product states.

\begin{figure}
    \centering
    \includegraphics[width=\columnwidth]{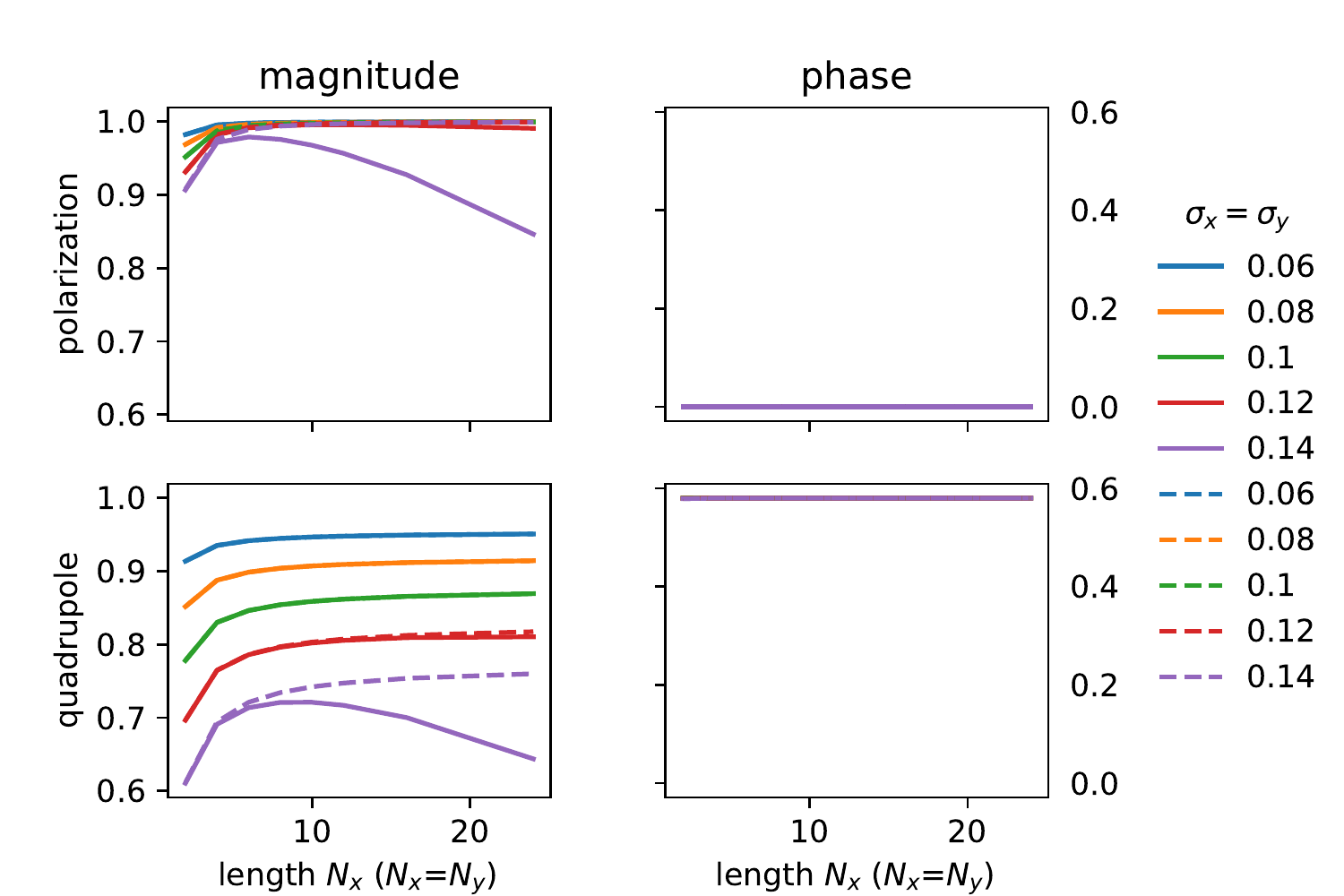}
    \caption{Comparison of determinant (solid lines) and product states (dashed lines) of Gaussian orbitals as a function of system size, keeping $N_x=N_y$ (always even) and $\sigma_x=\sigma_y$. For all these calculations, we fixed $x_0=y_0=0.2$. Top left: magnitude of the polarization operator goes to zero for the determinant with larger values of $\sigma_x$, but goes to one for the product state, regardless of $\sigma_x$. Top right: phase of the polarization operator is always zero. Bottom left: Magnitude of the quadrupole operator plateaus in the infinite limit to a value depending on $\sigma_x$ for both determinant and product states. For large enough $\sigma$, the magnitude from the determinant state tends toward zero in the infinite limit. Bottom right: The phase of the quadrupole operator is consistent for all calculations. For odd values of $N_x=N_y$ (not shown), the quadrupole phase is shifted by $0.5$.}
    \label{fig:gaussian_determinant}
\end{figure}

We confirm that the determinant state yields similar results to the product state except when the Gaussian spreads $\sigma_x,\sigma_y$ grow so large that determinant state exhibits a transition to a conducting state (i.e., the magnitude of the polarization operator goes to zero in the infinite limit), while the product state remains insulating (Fig. \ref{fig:gaussian_determinant}).
For smaller values of $\sigma_x, \sigma_y$, the phases and magnitudes of the product state agree with the determinant state.
We confirm that the phase of the polarization operator expectation value in our states is always zero, so the quadrupole is well-defined in principle.
We also confirm that the magnitude of the quadrupole operator remains fixed when we vary $\sigma=\sigma_x=\sigma_y$ and even values of $N_x=N_y$, and is the same for the determinant and product states.
When $\sigma$ increases, even when the magnitude of the polarization goes to one, the limit of the quadrupole magnitude decreases (it does not always approach one, even when finite).
Beyond a certain value of $\sigma$, the quadrupole magnitude goes to zero.
Based on our numerics, the insulating-conducting transition of the determinant state seems to coincide with the transition where the quadrupole magnitude goes to zero.

Our simple model of Gaussian orbitals can only go so far in explaining the results from our tightbinding calculations; however, we have shown that even with very simple wave functions, the quadrupole magnitude is suppressed by dipole fluctuations, and does not generally tend to one in the thermodynamic limit. We expect that our results would closely match the tightbinding results if we accounted for dipole fluctuations across the periodic boundary conditions.

\bibliography{quad_references}

\end{document}